\documentclass[aps,pre,preprint,showpacs,bibnotes]{revtex4}

\usepackage{amsmath,amsfonts,graphics,epstopdf}

\begin{document}

\title{The Wigner distribution and 2D classical maps} 

\author{Jamal Sakhr}
\affiliation{Department of Physics and Astronomy, University of Western Ontario, London, Ontario N6A 3K7 Canada}

\date{\today}

\begin{abstract}
The Wigner spacing distribution has a long and illustrious history in nuclear physics and in the quantum mechanics of classically chaotic systems. In this paper, a novel connection between the Wigner distribution and 2D \emph{classical} mechanics is introduced. Based on a well-known correspondence between the Wigner distribution and the 2D Poisson point process, the hypothesis that typical pseudo-trajectories of a 2D ergodic map have a Wignerian nearest-neighbor spacing distribution (NNSD) is put forward and numerically tested. The standard Euclidean metric is used to compute the interpoint spacings. In all test cases, the hypothesis is upheld, and the range of validity of the hypothesis appears to be robust in the sense that it is not affected by the presence or absence of: (i) mixing; (ii) time-reversal symmetry; and/or (iii) dissipation.  
\end{abstract}


\maketitle

\section{Introduction} 

Random matrix theory (RMT) \cite{Porter,Brody81,Mehta}, which was originally formulated in the 1960s to better understand nuclear spectra, has more recently emerged as a fundamental theoretical tool for understanding the spectra of quantum systems whose classical analogs are chaotic (see Refs.~\cite{gutz,Bohigas91,Guhr,Stockmann,Haake,Reichl} for examples and discussions). The most basic mathematical object of the theory is the probability density $P(S)$ of the spacing $S$ between adjacent energy levels. The aforementioned probability distribution $P(S)$, commonly referred to as the nearest-neighbor spacing distribution (NNSD), has been computed for a vast number of experimental and theoretical systems (mostly bounded conservative systems), and its overall shape varies depending on: (i) the presence or absence of time-reversal symmetry; and (ii) the degree of chaoticity in the classical dynamics. When there is time-reversal symmetry and the classical dynamics are strongly-chaotic, the sample NNSD is typically well modeled by the Mehta-Gaudin distribution, which is the theoretical NNSD for the eigenvalues of a random matrix chosen from the so-called Gaussian orthogonal ensemble (GOE). The latter distribution, denoted here by $P_{GOE}(S)$, cannot be expressed in closed-form (i.e., in terms of a finite number of elementary functions). An excellent analytical approximation to $P_{GOE}(S)$ is however given by the Wigner distribution \cite{Haake,Mehta}
\begin{equation}\label{WigdisRMT}
P_W(S)={\pi\over2}S\exp\left(-{\pi\over4}S^2\right), 
\end{equation}
which happens to be the theoretical NNSD for eigenvalues chosen from a Gaussian ensemble of real symmetric $2\times2$ random matrices.

As alluded to above, the quantum eigenvalue spectra of many strongly-chaotic conservative systems have been observed to possess ``Wigner-\emph{like}'' (or ``GOE-\emph{like}'') NNSDs. In view of this fact, a ``Wigner-\emph{like}'' NNSD is now widely regarded to be a ``generic'' property of time-reversal-invariant (TRI) quantum systems having strongly-chaotic classical limits \footnote{There are cases for which the NNSD is qualitatively different from $P_W(S)$ (for classic examples, see Refs.~\cite{Delande1,Delande2}). Such cases are deemed to be ``special'' or ``non-generic''.}. The underlying reasons for this observed statistical behavior nevertheless remain elusive. It is \emph{presumed} that, in ``generic'' TRI systems, the Wignerian shape of the NNSD (a property of the quantum eigenvalues) derives solely from the chaoticity of the classical dynamics (a property of the classical trajectories). It is however not fundamentally understood \emph{how} classical chaos is itself responsible for producing the observed Wignerian shape of the energy-level spacing distribution (a point that is often glossed over in the ``quantum chaos'' literature).  

While there have been significant advancements in understanding the observed agreement with other 
predictions of the random matrix model, such as the spectral rigidity \cite{mbdel3} and the two-point correlation function \cite{BogoKeat96,Sieber01,Sieber02,Huesler1,Huesler2,Huesler3,Keating07}, there has been comparatively little material progress in understanding the observed agreement (most often semi-quantitative) with the eigenvalue spacing statistics of the classical random matrix ensembles. The nearest-neighbor spacing distribution is particularly challenging mathematically. An analytical means of understanding the spacing statistics of quantum chaotic systems, based on semiclassical periodic orbit theory, has been attempted in Ref.~\cite{Keppeler2001}. However, due to the asymptotic techniques employed therein, the semiclassical formulas for the $k$th-nearest-neighbor spacing distributions $P(S;k)$ are good approximations to the corresponding distributions obtained from RMT only for large values of $S$ and $k$; for small $k$ and in particular for $k=1$ the obtained semiclassical formulas do not reproduce the well-known RMT results. In short, a clear-cut theoretical justification for \emph{why} a Wigner-\emph{like} NNSD is a common ``quantum signature of chaos'' is still lacking.

Curiously, the Wigner distribution also appears in the 
seemingly unrelated subject of spatial point processes, and more specifically, in the context of the
homogeneous Poisson point process in $\mathbb{R}^2$ (henceforth denoted 
by $\mathbf{P}_2$): the Wigner distribution is the nearest-neighbor spacing distribution (NNSD) for $\mathbf{P}_2$ \cite{Haake,meandjohn}. To the author's knowledge, 
this result was first mentioned in a paper by Grobe, Haake, and Sommers \cite{Grobe88} published in 1988 
and was subsequently useful in understanding certain spectral fluctuation 
properties of quantum dissipative systems (see, for instance, Chapter 9 of Ref.~\cite{Haake}). 
The intent of this paper is to introduce a connection between $\mathbf{P}_2$ and two-dimensional (2D) ergodic maps and to thereby explicate the fundamental significance of $P_W(S)$ to 2D \emph{classical} mechanics.

\begin{figure}
\scalebox{0.493}{\includegraphics*{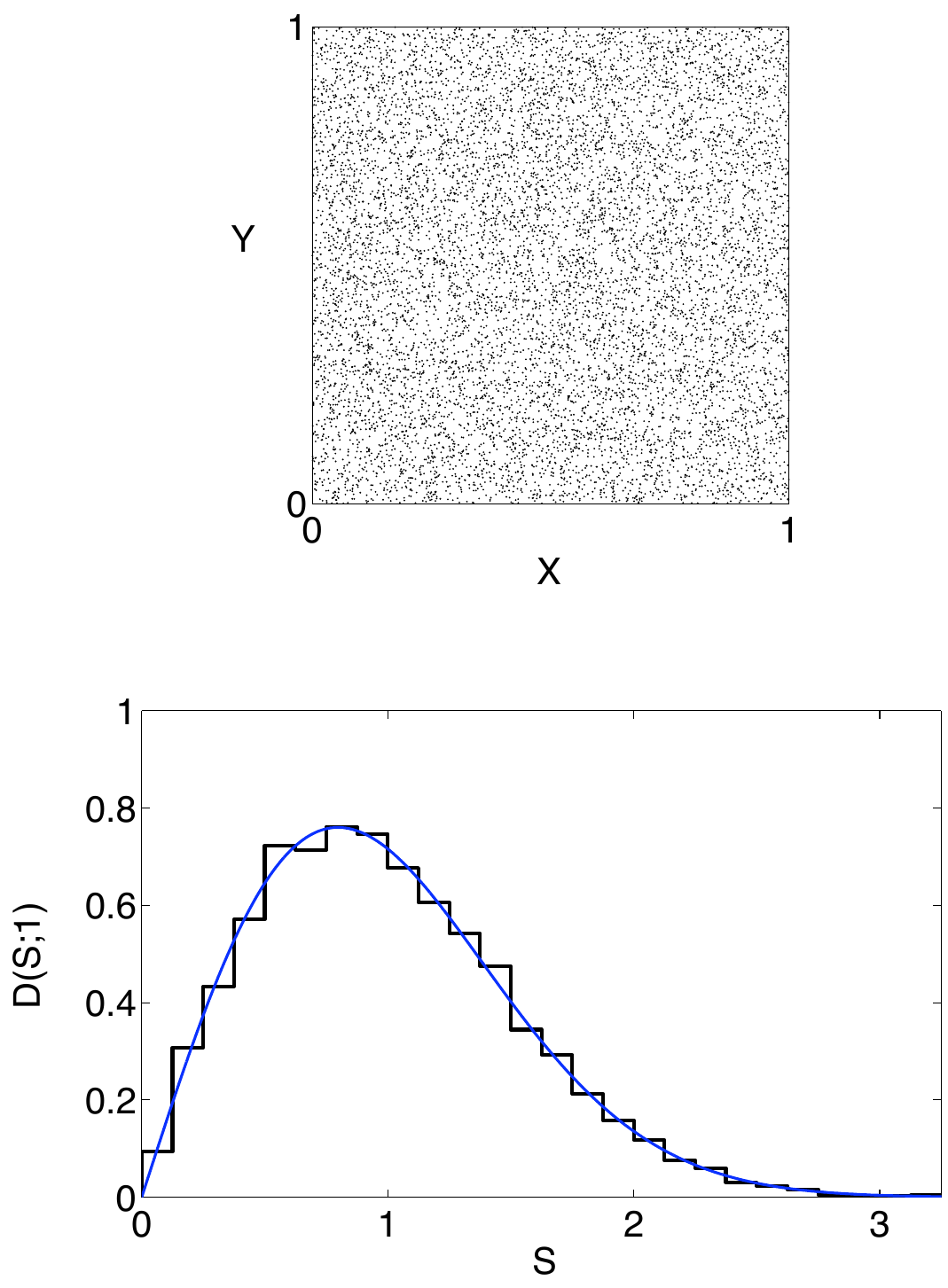}}
\caption{\label{P2andWig} (Top) A numerical realization of a two-dimensional homogeneous Poisson point process (of intensity 10000) consisting of $10000$ random points uniformly distributed in the unit square. (Bottom) Interpoint nearest-neighbor spacing density histogram for the point pattern shown above. The smooth curve is the Wigner distribution [Eq.~(\ref{WigdisRMT})].}
\end{figure}

\section{Ergodic Trajectories of 2D Classical Maps: The $\mathbf{P}_2$ Model}\label{theproposal} 

The link between $\mathbf{P}_2$ and 2D classical mechanics to be introduced below is based on a simple observation: Trajectories of discrete maps are discrete point sets. The most well-known example is the Poincar\'{e} return map. In numerical investigations of discrete maps, it is common to compute numerical trajectories (i.e., `pseudotrajectories') for a large number of initial conditions and to then generate a so-called phase portrait (a plot consisting of all computed pseudotrajectories overlaid in a single window whose boundary defines the phase space of the map). For both symplectic and dissipative maps, the signature phase portrait of an ergodic map is an apparently random scatter of points (often generated from \emph{one} initial condition). This signature point pattern has become ubiquitous in numerical investigations, especially in numerical Poincar\'{e} surface-of-section (SOS) computations for chaotic Hamiltonian systems, and the understanding in that context is that the apparently random scatter of points is a numerical rendering of how an ergodic (or nearly ergodic) orbit explores the interior of the energy shell in the full phase space. In the specific context of 2D maps, such point patterns are \emph{visually} indistinguishable from any numerical realization of $\mathbf{P}_2$ (see, for example, the top panel of Fig.~\ref{P2andWig}). This begs the question: \emph{For a 2D ergodic map, do the interpoint nearest-neighbor spacings of any typical pseudotrajectory of that map have a Wignerian distribution?} 

Intuitively, the affirmative answer is correct. Any typical pseudotrajectory of a 2D ergodic map will densely and uniformly cover the entire phase space of that map (excluding perhaps a zero-measure subset of the phase space) \footnote{In this paper, the term ``ergodic'' should be understood as a shorthand for simply ergodic, which means that typical pseudotrajectories will (after sufficient time) cover the phase space uniformly. Non-simple ergodicity arising from dense but non-uniform coverage of the phase space occurs in many well-known 2D maps (e.g., the Sinai map), but the treatment of this type of ergodicity requires certain refinements that lie beyond the scope of the present paper.}. The interpoint spacing statistics of such a pseudotrajectory should therefore be consistent with the spacing statistics theoretically predicted for $\mathbf{P}_2$. In particular, the NNSD of such pseudotrajectories should be consistent with the Wigner distribution. These arguments should hold regardless of the presence or absence of: (i) mixing; (ii) time-reversal symmetry; and (iii) dissipation. Furthermore, pseudotrajectories need not necessarily cover densely all of the available phase space in order for their NNSDs to be Wignerian. Pseudotrajectories evolving ergodically in any positive-measure subset of the full phase space should also possess NNSDs consistent with the Wigner distribution. The preceding claims naturally require verification and indeed the intent in the following sections of the paper is to validate these claims numerically. 

\section{Generic Examples of 2D Ergodic Maps}\label{modexamples}

\subsection{Example 1: Reversible Symplectic Anosov Map}\label{modexample1} 

\begin{figure}
\scalebox{0.163}{\includegraphics*{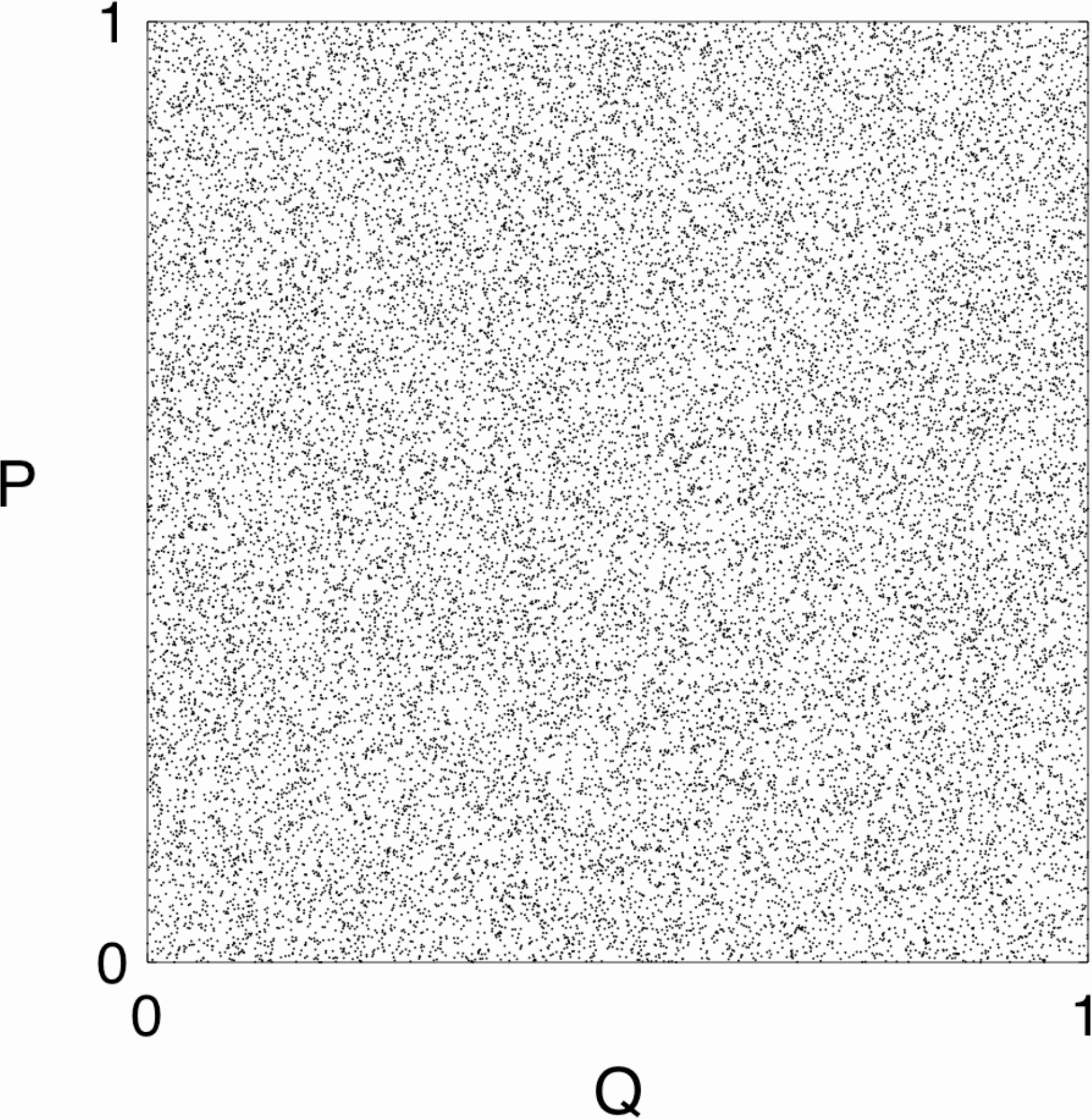}} \hspace*{1cm}
\scalebox{0.373}{\includegraphics*{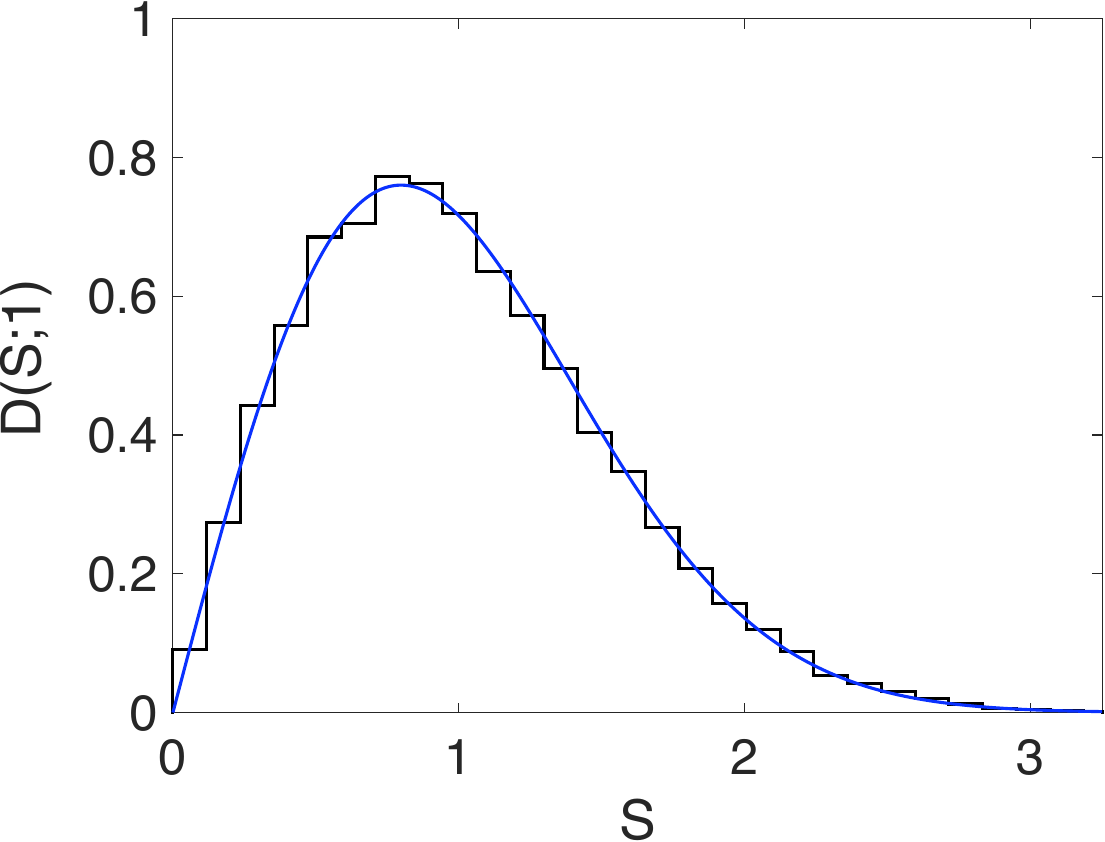}} \par \vspace*{0.75cm}
\scalebox{0.163}{\includegraphics*{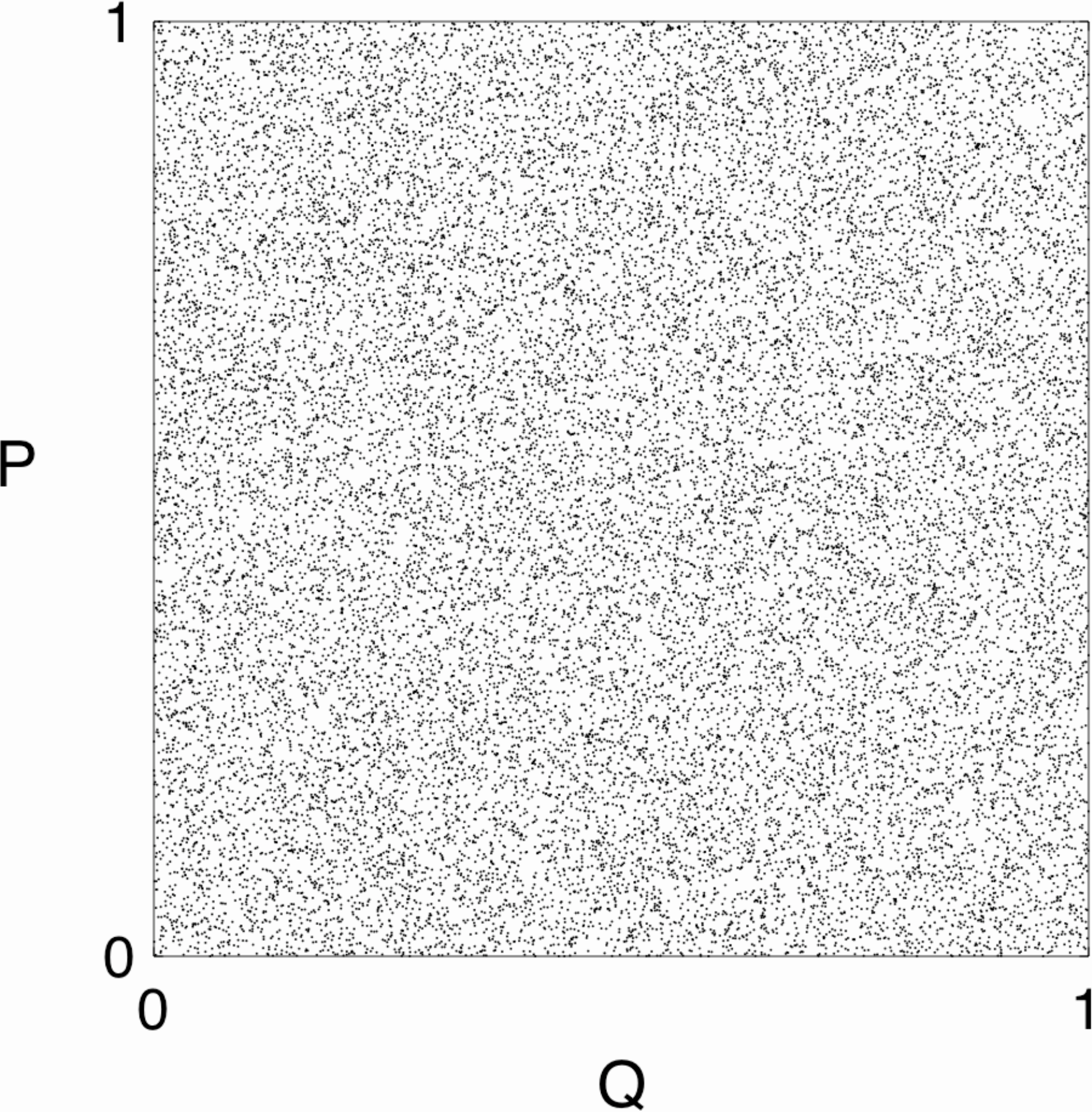}} \hspace*{1cm}
\scalebox{0.373}{\includegraphics*{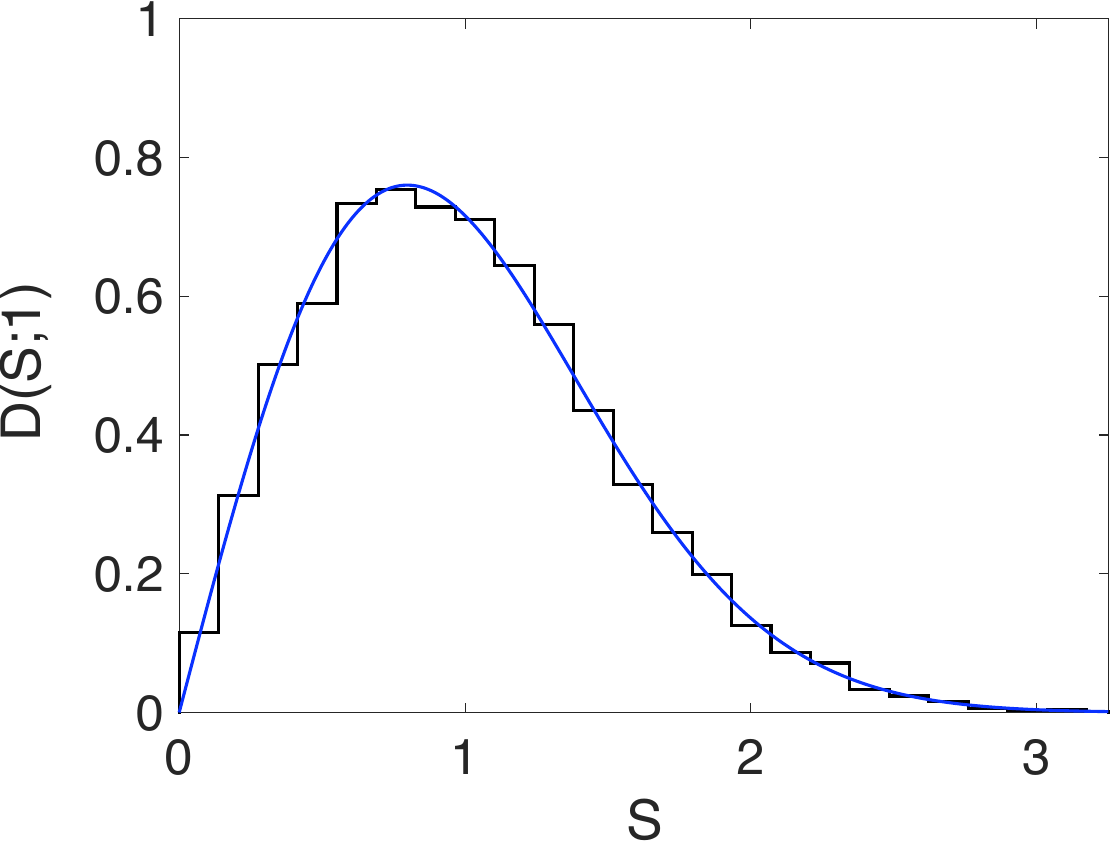}}
\caption{\label{catTRIeg} (Top Left) A typical pseudotrajectory of the unperturbed cat map defined in Eq.~(\ref{unperturbedcatTRI}). The pseudotrajectory was evolved from the initial point $(q_1=(\sqrt{2}-1)/3,p_1=(\sqrt{3}-1)/3)$, and the map was iterated 25000 times. (Top Right) Nearest-neighbor spacing density histogram for the pseudotrajectory shown to the left. The smooth curve is the Wigner distribution [Eq.~(\ref{WigdisRMT})]. (Bottom Left) A typical pseudotrajectory of the perturbed cat map (\ref{perturbedcatTRI}) with $K=0.3$. The pseudotrajectory was evolved from the initial point $(q_1=(\sqrt{2}-1)/3,p_1=(\sqrt{3}-1)/3)$, and the map was iterated 25000 times. (Bottom Right) Nearest-neighbor spacing density histogram for the pseudotrajectory shown to the left. The smooth curve is the Wigner distribution [Eq.~(\ref{WigdisRMT})].}
\end{figure}

The cat map 
\begin{eqnarray}\label{unperturbedcatTRI}
\left(\begin{array}{c} q_{n+1} \\ p_{n+1}\end{array} \right)=\left(\begin{array}{cc} 2 & 1 \\ 3 & 2\end{array}\right)\left(\begin{array}{c} q_{n} \\ p_{n}\end{array} \right),~\text{mod~1} 
\end{eqnarray}
is area-preserving, ergodic and mixing (in fact hyperbolic), and time-reversal invariant \cite{Keating98}. In studying `typical' (non-periodic) pseudotrajectories of map (\ref{unperturbedcatTRI}), the set of initial conditions consisting of all pairs of rational numbers chosen from the unit interval is excluded. This is a set of measure zero in $\mathbb{R}^2$, and thus almost all initial conditions chosen from $[0,1]\times[0,1]$ produce long pseudotrajectories suitable for analysis. A typical pseudotrajectory of map (\ref{unperturbedcatTRI}) is shown in the top left panel of Figure \ref{catTRIeg}. The task now is to analyze the spacings between the points of this pseudotrajectory.  

The distance between two points $\xi_i=(q_i,p_i)$ and $\xi_j=(q_j,p_j)$ in phase space is defined here using the two-dimensional Euclidean metric: $\Delta(\xi_i,\xi_j)=\sqrt{(q_i-q_j)^2+(p_i-p_j)^2}$. 
The distance between a given point $\xi_i$ and its nearest neighbor is then defined by 
$d^{(1)}_{i} = \text{min} \left\{\Delta\left(\xi_i,\xi_j\right) 
: i,j=1,\ldots,N~(j \neq i)\right\}$, and similarly the distance between 
$\xi_i$ and its furthest neighbor is defined by 
$d^{(N)}_{i} = \text{max} \left\{\Delta\left(\xi_i,\xi_j\right) 
: i,j=1,\ldots,N~(j \neq i)\right\}$. If, for each given point  
$\xi_i$, the spacings $\left\{\Delta\left(\xi_i,\xi_j\right) 
: i,j=1,\ldots,N~(j \neq i)\right\}$ are sorted by size (in ascending order), then the 
$k$th-nearest-neighbor spacing is the $k$th element of the set
$\{d^{(1)}_i,d^{(2)}_i,\ldots,d^{(k)}_i,\ldots,d^{(N)}_i\}$. For numerical comparisons, the sample $k$th-NNSD is defined in terms of the scaled spacings $S^{(k)}_i = d^{(k)}_i / \bar{d}^{(k)}$, 
where $\bar{d}^{(k)}=(1/N)\sum_{i=1}^N d^{(k)}_i$, is the 
mean $k$th-nearest-neighbor spacing. The $k$th-nearest-neighbor spacing density histogram is then constructed by binning all $N$ values of $S^{(k)}_i$ and normalizing the area under the histogram to unity. 

For the pseudotrajectory shown in the top left panel of Figure \ref{catTRIeg}, the density histogram of the (scaled) nearest-neighbor spacings is shown in the top right panel of Figure \ref{catTRIeg}. The latter is clearly in accord with the Wigner distribution [Eq.~(\ref{WigdisRMT})].

Small nonlinear perturbations of map (\ref{unperturbedcatTRI}) that preserve the time-reversal symmetry do not alter the above result. Consider, for example, the perturbed cat map \cite{backer03}
\begin{eqnarray}\label{perturbedcatTRI}
\left(\begin{array}{c} q_{n+1} \\ p_{n+1}\end{array} \right)=\left(\begin{array}{cc} 2 & 1 \\ 3 & 2\end{array}\right)\left(\begin{array}{c} q_{n} \\ p_{n}\end{array} \right) + {K\over2\pi}\left(\begin{array}{c} 1 \\ 2\end{array} \right)\cos(2\pi q_n),~\text{mod~1} 
\end{eqnarray}
where $K$ is a free perturbation parameter. Map (\ref{perturbedcatTRI}) is symplectic and time-reversal-invariant \cite{OZA1}. It is also hyperbolic when $K\le\left(\sqrt{3}-1\right)/\sqrt{5}$ \cite{backer03}. A typical pseudotrajectory of map (\ref{perturbedcatTRI}) with $K=0.3$ is shown in the bottom left panel of Figure \ref{catTRIeg} and the density histogram of the (scaled) nearest-neighbor spacings is shown in the bottom right panel of Figure \ref{catTRIeg}. The latter is again consistent with the Wigner distribution.   
 
\subsection{Example 2: Reversible Symplectic Map that is Ergodic but not Mixing}\label{modexample2}

\begin{figure}
\scalebox{0.323}{\includegraphics*{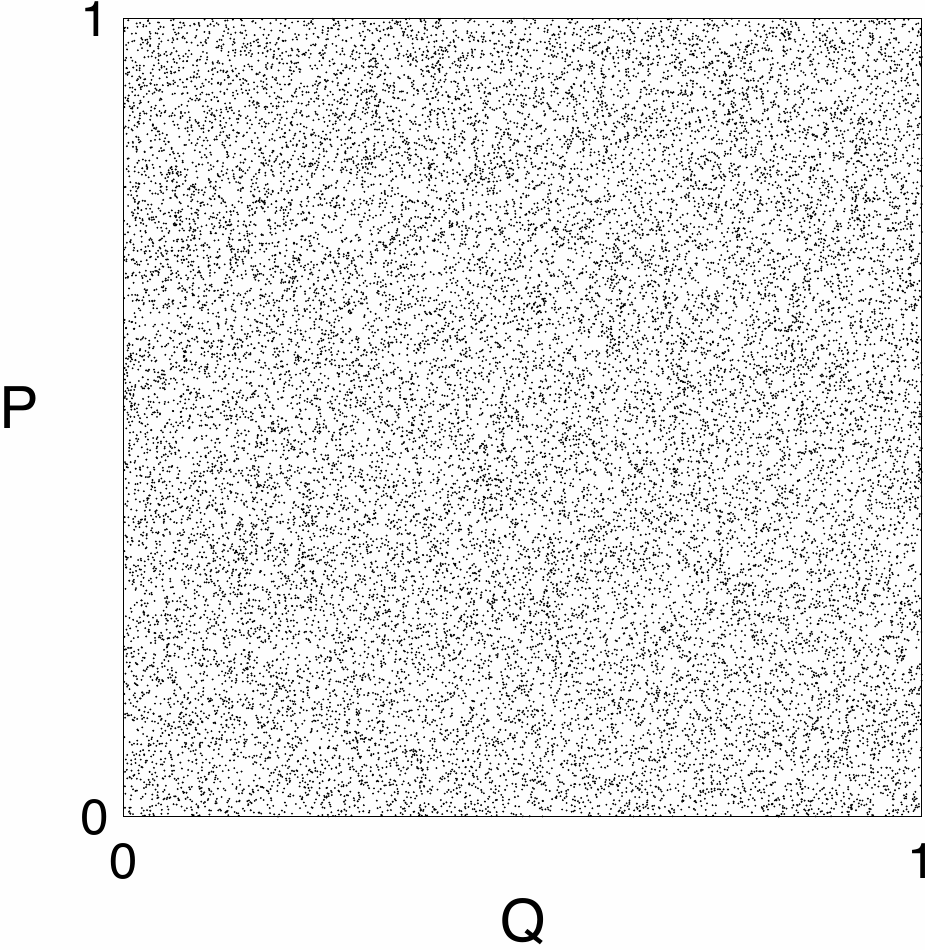}} \hspace*{1cm}
\scalebox{0.373}{\includegraphics*{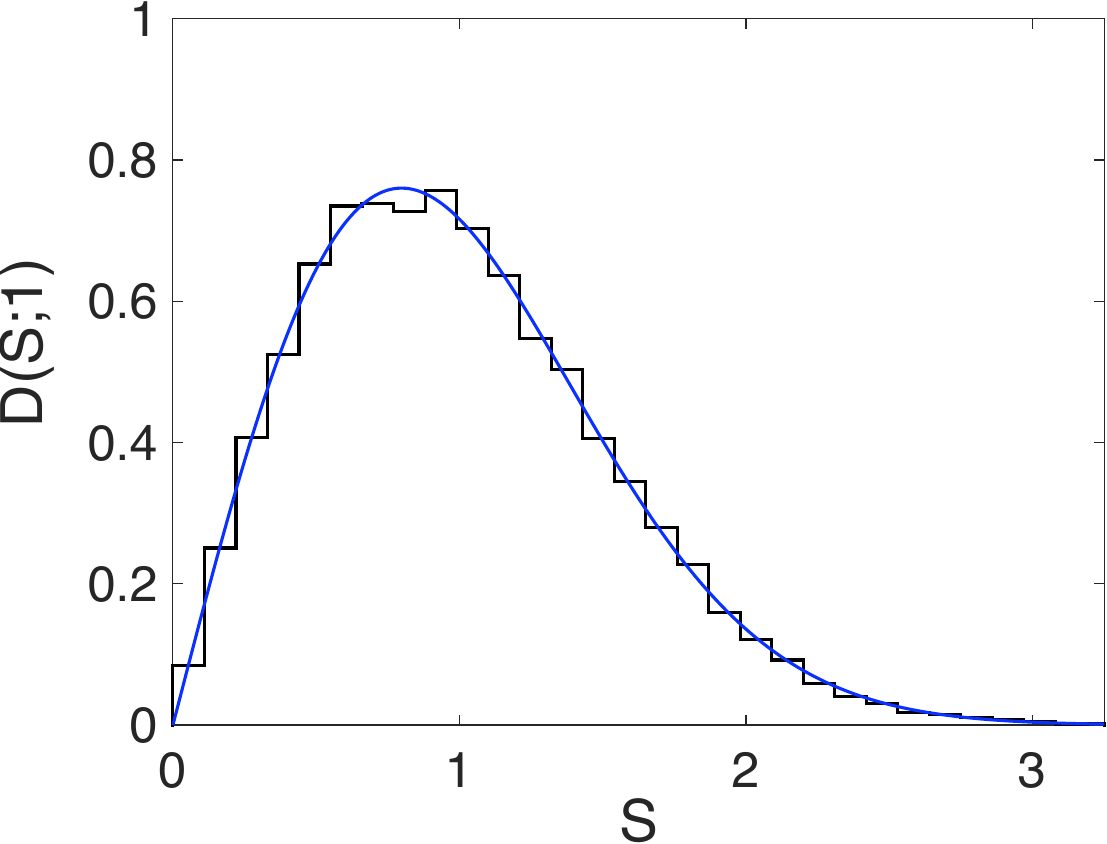}}
\caption{\label{skewEG} (Left) A typical pseudotrajectory of the non-mixing ergodic map defined in Eq.~(\ref{NCmap}) when $\alpha=\ln(2)$. In this instance, the pseudotrajectory was evolved from the initial point $(p_1=1/\sqrt{11},q_1=1/\sqrt{11})$, and the map was iterated 25000 times. (Right) Nearest-neighbor spacing density histogram for the pseudotrajectory shown to the left. The smooth curve is the Wigner distribution [Eq.~(\ref{WigdisRMT})].}
\end{figure}

Consider next the following 2D map that describes skew translations (or rotations) on the unit two-torus:
\begin{eqnarray}\label{NCmap}
\left(\begin{array}{c} q_{n+1} \\ p_{n+1} \end{array} \right)=\left(\begin{array}{cc} 1 & 0 \\ 2 & 1\end{array}\right)\left(\begin{array}{c} q_{n} \\ p_{n} \end{array} \right)+\alpha\left(\begin{array}{c} 1 \\ 0\end{array} \right),~\text{mod~1}.
\end{eqnarray}
When the real parameter $\alpha$ is an irrational number, map (\ref{NCmap}) is (uniquely) ergodic and \emph{not} mixing \cite{Cornfield,Fguy}. Map (\ref{NCmap}) is also time-reversal invariant  \cite{spinny02}. The left panel of Figure \ref{skewEG} shows a typical pseudotrajectory of map (\ref{NCmap}) when $\alpha=\ln(2)$, and the density histogram of the (scaled) nearest-neighbor spacings is shown in the right panel of Figure \ref{skewEG}. The latter is again consistent with the Wigner distribution. Incidentally, the nearest-neighbor spacing distribution of the eigenphases that come from quantizing map (\ref{NCmap}) is not well-defined in the semiclassical limit \cite{Backerskew}. 

\subsection{Example 3: Reversible Symplectic Map with an Ergodic Component}\label{divedeg}

\begin{figure}
\scalebox{0.183}{\includegraphics*{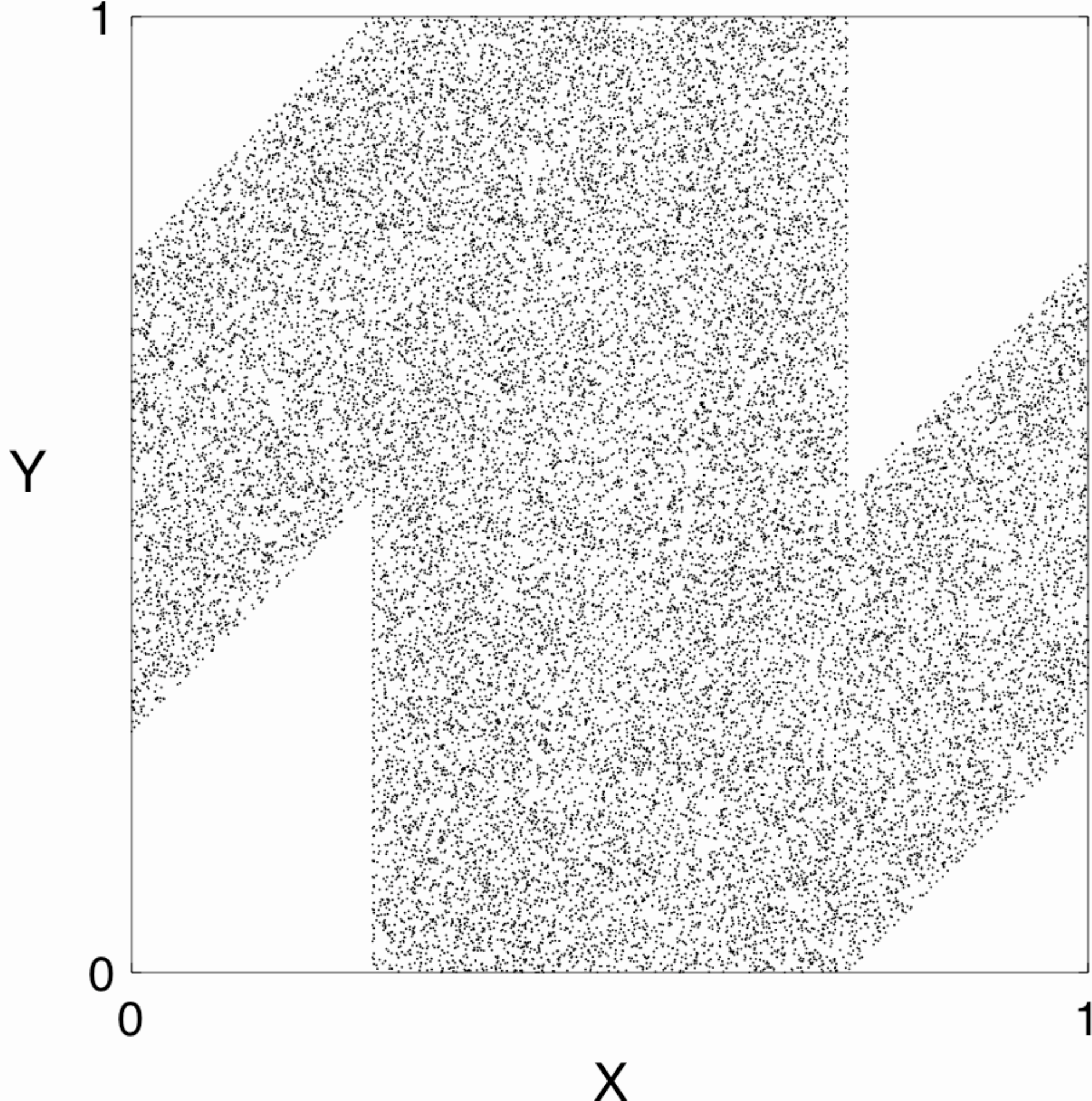}}  \hspace*{1cm}
\scalebox{0.313}{\includegraphics*{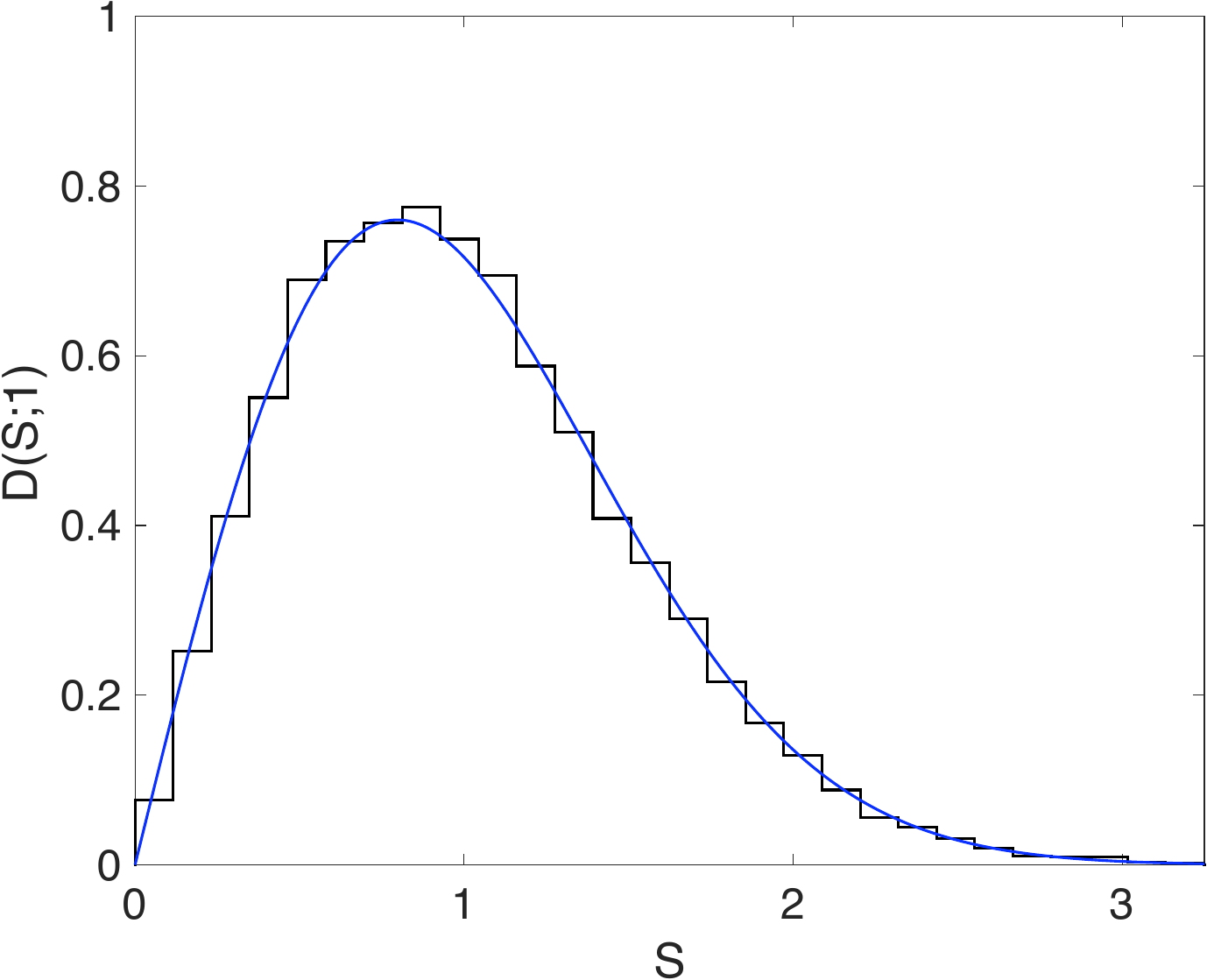}}
\caption{\label{dividedEG} (Left) A typical pseudotrajectory of map (\ref{genmap}) with parameter $K=2$. The pseudotrajectory was evolved from the initial point $(x_1=1/\sqrt{5},y_1=1/\sqrt{5})$, and the map was iterated $25000$ times. (Right) Nearest-neighbor spacing density histogram for the pseudotrajectory shown to the left. The smooth curve is the Wigner distribution [Eq.~(\ref{WigdisRMT})].}
\end{figure}

2D area-preserving maps of the form 
\begin{subequations}\label{genmap}
\begin{equation}\label{genmapp1}
y_{n+1}=y_n+K f(x_n),~\text{mod}~1,
\end{equation}
\begin{equation}\label{genmapp2}
x_{n+1}=x_n+y_{n+1},~\text{mod}~1,
\end{equation}
where $K$ is a free parameter, and $f(x_n)$ is some prescribed function, generally possess a mixed phase space consisting of commingled regular and chaotic regions \cite{Zas98}. (The ``standard map'' obtained by prescribing $f(x_n)={1\over2\pi}\sin(2\pi x_n)$ is the most well-known and well-studied member of this family of maps.) The form and distribution of the regions generally depend on both the function $f$ and the parameter $K$. When the function $f(x_n)$ is a piece-wise linear function of the interval $x_n\in[0,1]$, the phase space of the above map is mixed but can be sharply divided in the sense that regular and chaotic regions are separated by a simple curve \cite{Wkowski81}. For example, when 
\begin{eqnarray}\label{PLfunc}
f(x_n)=\left \{ \begin{array}{lr}
           -x_n & ~\text{if}~x_n\in\left[0,{1\over4}\right) \\ 
           -{1\over2}+x_n & ~\text{if}~x_n\in\left[{1\over4},{3\over4}\right) \\
           1-x_n & ~\text{if}~x_n\in\left[{3\over4},1\right] \\
           \end{array} \right.,
\end{eqnarray}
\end{subequations}
the phase space consists of one regular region and one ergodic region separated only by the border of the regular region, which in this case is non-hierarchical \cite{Lee89}. In the present context, the most important feature of map (\ref{genmap}) is that its phase space has an ergodic region (i.e., there exists a positive-measure subset of the full phase space wherein the dynamics of map (\ref{genmap}) is ergodic). As argued in Sec.~\ref{theproposal}, the interpoint NNSD of any typical pseudotrajectory initiated in the ergodic region should be consistent with the Wigner distribution. One such pseudotrajectory of map (\ref{genmap}) with $K=2$ is shown in the left panel of Figure \ref{dividedEG}, and the density histogram of the (scaled) nearest-neighbor spacings is shown in the right panel of Figure \ref{dividedEG}. Although the goodness-of-fit of the Wigner distribution appears to be lower here than in the previous examples, the Kolmogorov-Smirnov (KS) test does not reject the Wigner distribution at the $99\%$ significance level (KS test statistic=0.0070, \emph{p}-value=0.1776) nor does the more sensitive Anderson-Darling (AD) test (AD test statistic=2.3712, \emph{p}-value=0.0579) at the same significance level. 

\subsection{Example 4: Non-Reversible Symplectic Anosov Map}\label{modexample3}

\begin{figure}
\scalebox{0.323}{\includegraphics*{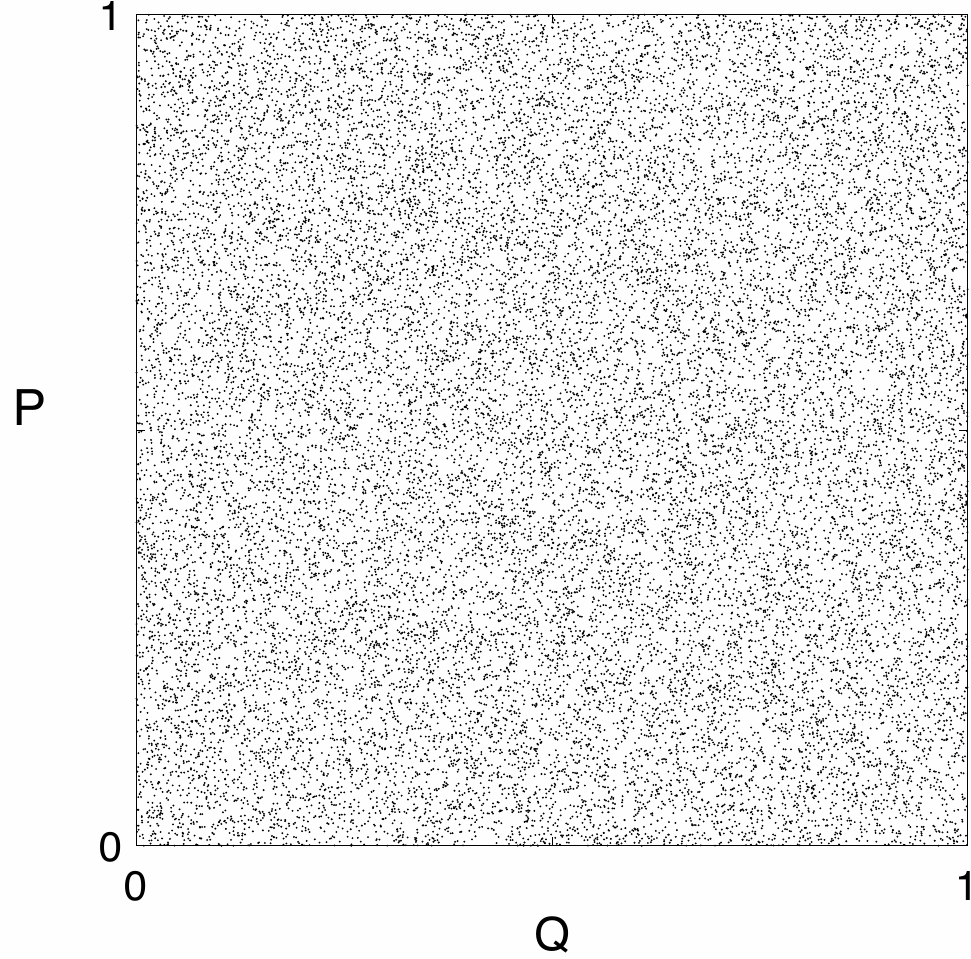}}  \hspace*{1cm}
\scalebox{0.373}{\includegraphics*{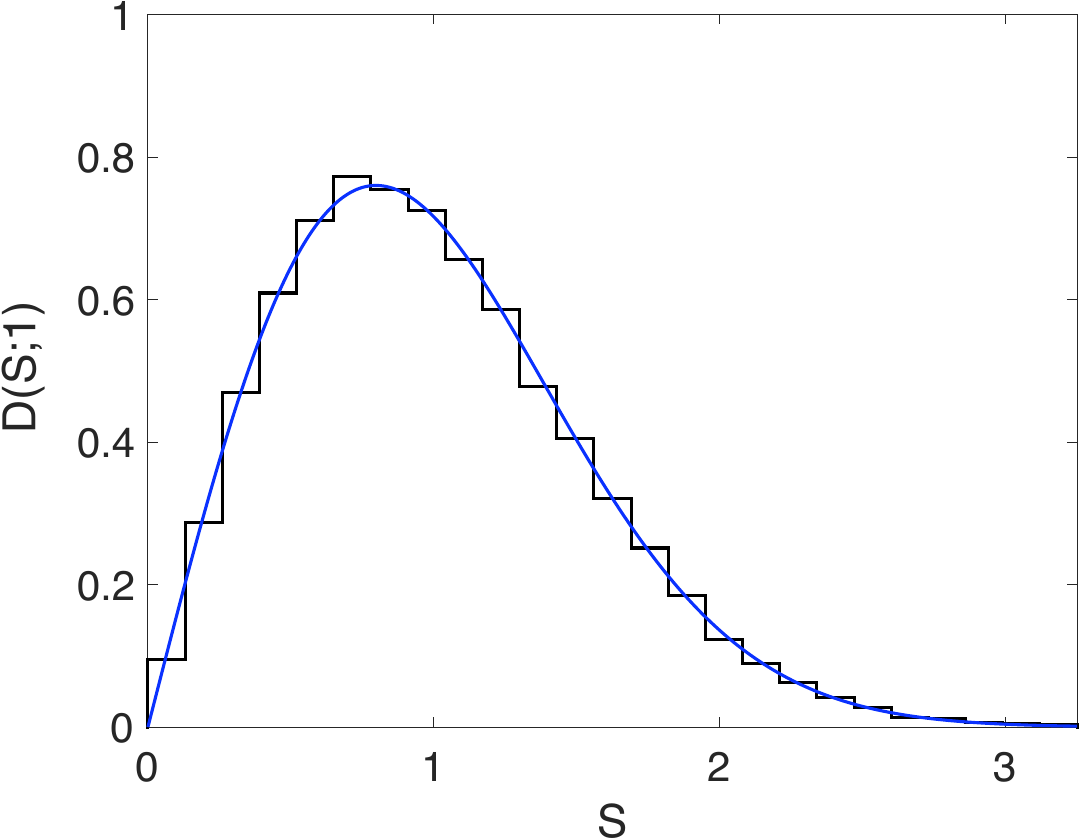}}  \par \vspace*{0.75cm}
\scalebox{0.323}{\includegraphics*{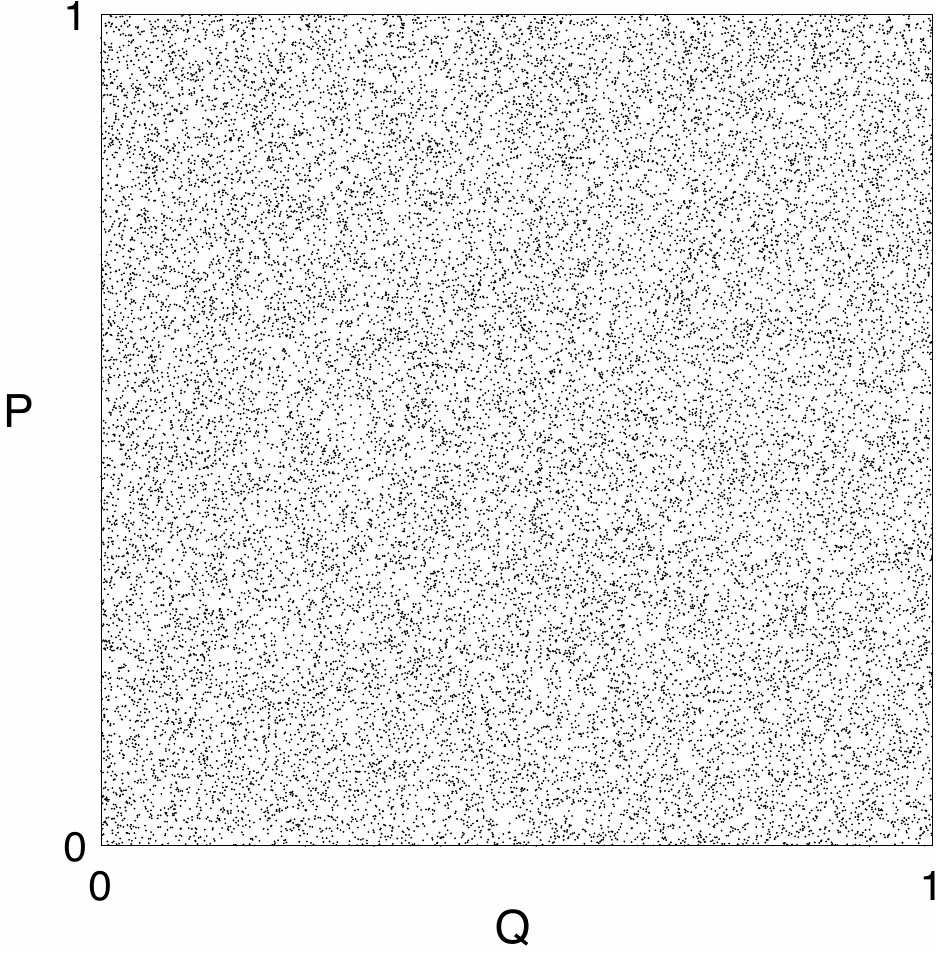}}  \hspace*{1cm}
\scalebox{0.373}{\includegraphics*{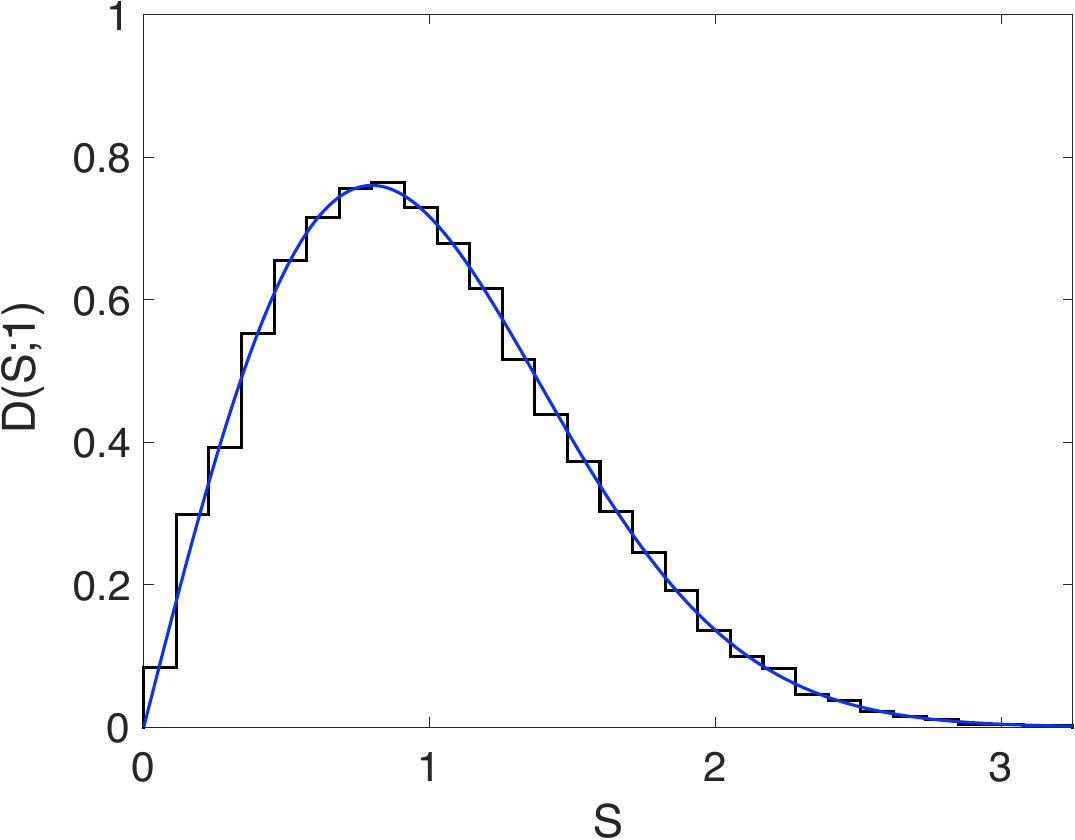}}
\caption{\label{unpertcat1EG} (Top Left) A typical pseudotrajectory of the unperturbed cat map defined in Eq.~(\ref{unperturbedcat1}). The pseudotrajectory was evolved from the initial point $(q_1=1/\sqrt{11},p_1=1/\sqrt{11})$, and the map was iterated 25000 times. (Top Right) Nearest-neighbor spacing density histogram for the pseudotrajectory shown to the left. The smooth curve is the Wigner distribution [Eq.~(\ref{WigdisRMT})]. (Bottom Left) A typical pseudotrajectory of the perturbed cat map defined in Eq.~(\ref{prtrbdmapTRSB}) with parameters $k_p=k_q=0.011$. The pseudotrajectory was evolved from the initial point $(q_1=(\sqrt{5}-1)/2,p_1=(\sqrt{5}-1)/3)$, and the map was iterated 25000 times. (Bottom Right) Nearest-neighbor spacing density histogram for the pseudotrajectory shown to the left. The smooth curve is the Wigner distribution [Eq.~(\ref{WigdisRMT})].}
\end{figure}

The area-preserving hyperbolic cat map 
\begin{eqnarray}\label{unperturbedcat1}
\left(\begin{array}{c} q_{n+1} \\ p_{n+1}\end{array} \right)=\left(\begin{array}{cc} 4 & 9 \\ 7 & 16\end{array}\right)\left(\begin{array}{c} q_{n} \\ p_{n}\end{array} \right),~\text{mod~1} 
\end{eqnarray}
is not time-reversal invariant \cite{Roberts97}. (In fact, the only symmetry possessed by map (\ref{unperturbedcat1}) is parity.) The top left panel of Figure \ref{unpertcat1EG} shows a typical pseudotrajectory of map (\ref{unperturbedcat1}) and the density histogram of the (scaled) \emph{nearest}-neighbor spacings, which is again consistent with the Wigner distribution, is shown in the top right panel of Figure \ref{unpertcat1EG}. 

From an RMT perspective, it is also interesting to consider small nonlinear perturbations of the above cat map that break so-called pseudosymmetries \cite{KM}. A simple example (following Ref.~\cite{KM}) is constructed by composing the unperturbed cat map given by Eq.~(\ref{unperturbedcat1}) with two shears: a momentum shear $P_p(q,p)=(q,p+k_pF(q))$, and a position shear $P_q(q,p)=(q+k_qG(p),p)$, where $k_p$, $k_q$ are free parameters that determine the strength of the perturbations, and $F(q)$ and $G(p)$ are periodic functions. If $A$ denotes the unperturbed cat map given by Eq.~(\ref{unperturbedcat1}), then the perturbed cat map is defined as follows:   
\begin{subequations}\label{prtrbdmapTRSB}
\begin{eqnarray}\label{prtrbdmapTRSB1}
\left(\begin{array}{c} q_{n+1} \\ p_{n+1}\end{array} \right)=\left(A \circ P_p \circ P_q\right)\left(\begin{array}{c} q_{n} \\ p_{n}\end{array} \right),~\text{mod~1}. 
\end{eqnarray}
More explicitly, the above map can be written as
\begin{equation}\label{prtrbdmapTRSB2}
\left(\begin{array}{c} q_{n+1} \\ p_{n+1}\end{array} \right)=\left(\begin{array}{cc} 4 & 9 \\ 7 & 16\end{array}\right)\left(\begin{array}{c} q_{n} \\ p_{n}\end{array} \right)+k_q\left(\begin{array}{c} 4 \\ 7\end{array} \right)G(p_n) + k_p\left(\begin{array}{c} 9 \\ 16\end{array} \right)F\left(q_n+k_qG(p_n)\right),~\text{mod~1}. 
\end{equation}
To break the parity symmetry of $A$ using the momentum shear $P_p$ (which is always invariant under time reversal), $F(q)$ should be an even function. To ensure that the perturbed map has no classical symmetries (and no pseudosymmetries when quantized), the functions $F(q)$ and $G(p)$ were chosen as follows: 
\begin{equation}
F(q)={1\over2\pi}\Big(\cos(2\pi q)-\cos(4\pi q)\Big),
\end{equation}
\begin{equation}
G(p)={1\over2\pi}\Big(\cos(4\pi p)-\cos(2\pi p)\Big).
\end{equation}
\end{subequations}
It is well-known that cat maps are structurally stable \cite{AA}, which means that, if the perturbation is sufficiently small, the perturbed cat map possesses the same dynamical properties as the unperturbed cat map; in other words, the hyperbolicity of the unperturbed map is preserved under sufficiently small perturbations. Map (\ref{prtrbdmapTRSB}) is thus symplectic and possesses no  symmetries (classical or otherwise), and (for small values of $k_p$ and $k_q$) the map is hyperbolic (and thus also ergodic). A typical pseudotrajectory of map (\ref{prtrbdmapTRSB}) with $k_p=k_q=0.011$ is shown in the bottom left panel of Figure \ref{unpertcat1EG} and the density histogram of the (scaled) nearest-neighbor spacings is shown in the bottom right panel of Figure \ref{unpertcat1EG}. The latter is again consistent with the Wigner distribution. The same result is obtained even when the parity symmetry of $A$ is not broken (i.e., when $F(q)$ is an odd function). 

\subsection{Example 5: Non-Reversible Dissipative Anosov Map}\label{modexample4}

\begin{figure}
\scalebox{0.373}{\includegraphics*{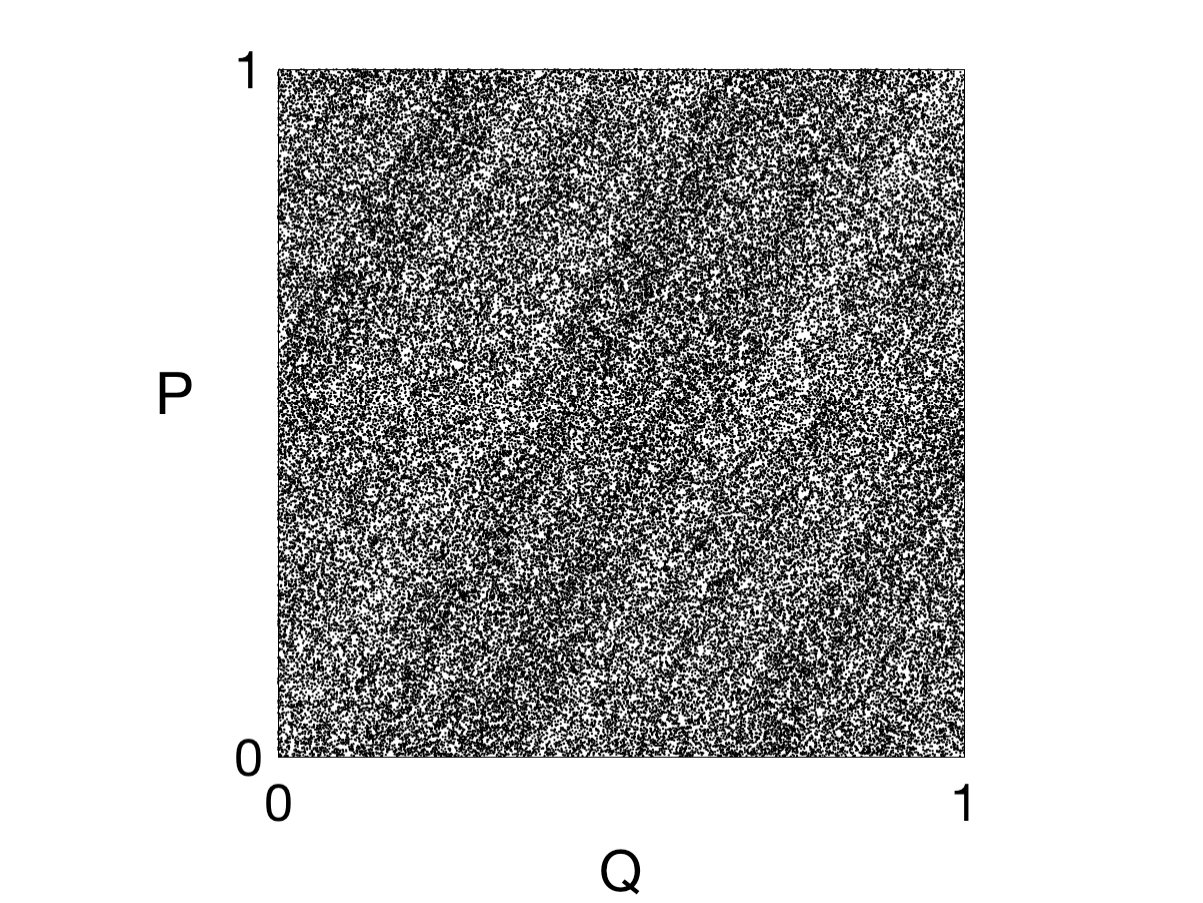}}
\scalebox{0.373}{\includegraphics*{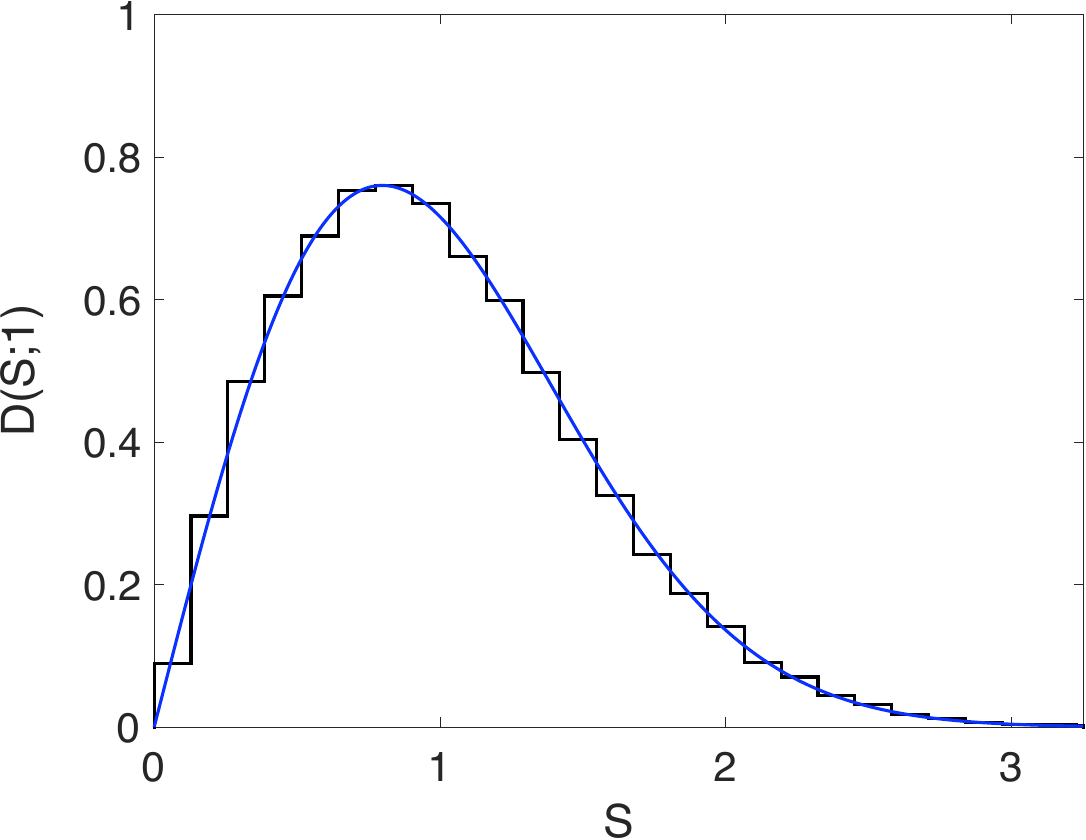}}
\caption{\label{pertcatdissEG} (Left) A typical pseudotrajectory of the dissipative cat map (\ref{prtrbdmapDISS}) with perturbation parameter $\varepsilon=0.033$. The pseudotrajectory was evolved from the initial point $(q_1=(\sqrt{5}-1)/5,p_1=(\sqrt{5}-1)/7)$, and the map was iterated $10^5$ times. (Right) Nearest-neighbor spacing density histogram for the pseudotrajectory shown to the left. The smooth curve is the Wigner distribution [Eq.~(\ref{WigdisRMT})].}
\end{figure}

Nonlinear perturbations of cat maps are generally not area-preserving. For example, the family of perturbed cat maps given by 
\begin{subequations}\label{prtrbdmapDISS}
\begin{eqnarray}\label{prtrbdmapDISS1}
\left(\begin{array}{c} q_{n+1} \\ p_{n+1}\end{array} \right)=\left(\begin{array}{cc} 1 & 1 \\ 1 & 2\end{array}\right)\left(\begin{array}{c} q_{n} \\ p_{n}\end{array} \right) + \varepsilon\left(\begin{array}{c} f_1(q_n,p_n) \\ f_2(q_n,p_n) \end{array} \right),~\text{mod~1} 
\end{eqnarray}
where $\varepsilon$ is a perturbation parameter, and $f_1(q_n,p_n)$ and $f_2(q_n,p_n)$ are periodic functions, is generally dissipative \cite{Gaspard05}. (Note that the unperturbed map is not reversible.) A simple but concrete example is furnished by taking (see Ref.~\cite{BGM00}) 
\begin{equation}
f_1(q_n,p_n)={1\over2\pi}\Big(\sin\left[2\pi\left(q_n+p_n\right)\right]+\sin\left(2\pi q_n\right)\Big),
\end{equation}
\begin{equation}
f_2(q_n,p_n)=0.
\end{equation}
\end{subequations}
Map (\ref{prtrbdmapDISS}) with $\varepsilon\neq0$ is dissipative with local volume contraction rate given by 
\begin{eqnarray}
\Lambda(q_n,p_n)\equiv\ln\big|\det{J(q_n,p_n)}\big|=\ln\Big |1+\varepsilon\Big\{\cos\left[2\pi\left(q_n+p_n\right)\right]+2\cos\left(2\pi q_n\right)\Big\}\Big |, 
\end{eqnarray}
where $J(q_n,p_n)$ is the Jacobian of map (\ref{prtrbdmapDISS}). For sufficiently small values of $\varepsilon$ (see Ref.~\cite{AA} for a precise criterion), map (\ref{prtrbdmapDISS}) is furthermore hyperbolic (and thus ergodic). A typical pseudotrajectory of map (\ref{prtrbdmapDISS}) with $\varepsilon=0.033$ is shown in the left panel of Figure \ref{pertcatdissEG} and the density histogram of the (scaled) nearest-neighbor spacings is shown in the right panel of Figure \ref{pertcatdissEG}. The latter is again consistent with the Wigner distribution. 

Note that longer pseudotrajectories ($N=10^5$ compared to the usual $N=25000$ used in all prior examples) were used in the analysis of this dissipative case. The reason is that the orbits of this dissipative map have a non-trivial spatial structure, and this structure takes some time to materialize. (The spatial structure of the orbits in the prior symplectic cases is comparatively trivial.) Shorter pseudotrajectories of this map (e.g., $N=25000$) also possess Wignerian NNSDs, but their statistical significance (as measured by goodness-of-fit tests) is lower. Longer pseudotrajectories (e.g., $N=10^6$) produce results very similar to those shown in Figure \ref{pertcatdissEG}, and their corresponding statistical significances are also similar. 

\subsection{Higher-Order Spacing Distributions}

\begin{figure}
\scalebox{0.523}{\includegraphics*{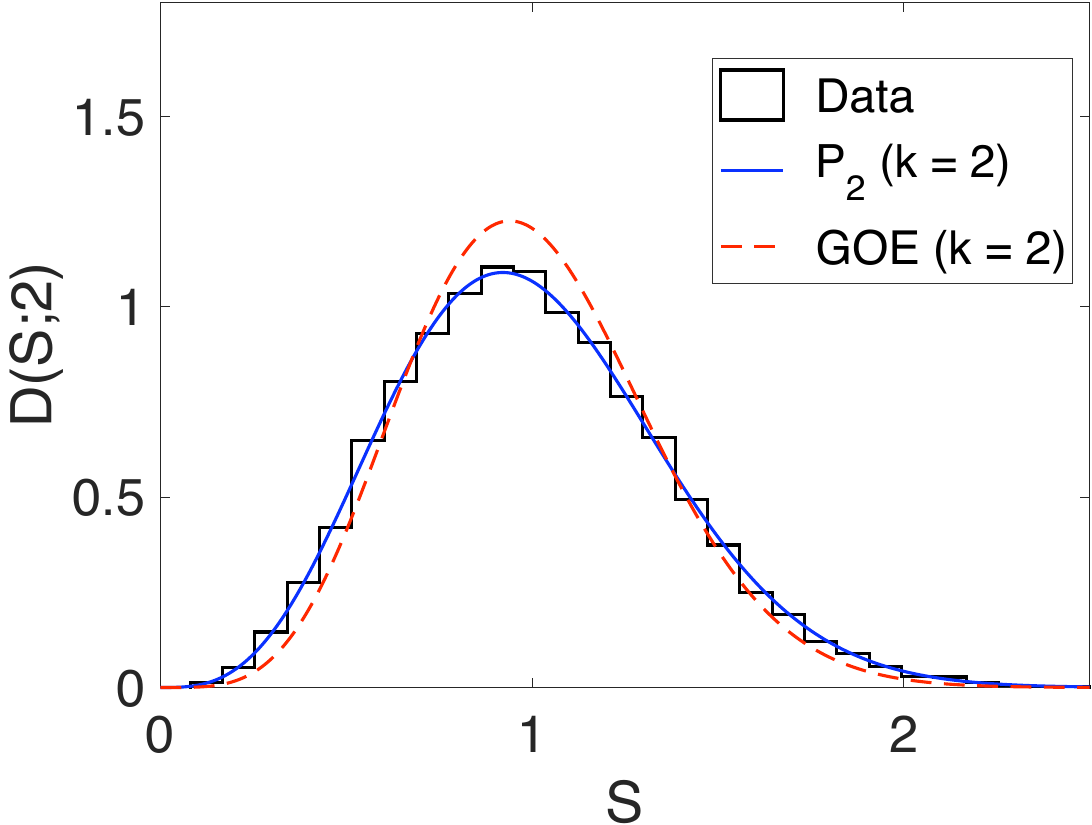}} \par \vspace*{0.5cm} 
\scalebox{0.523}{\includegraphics*{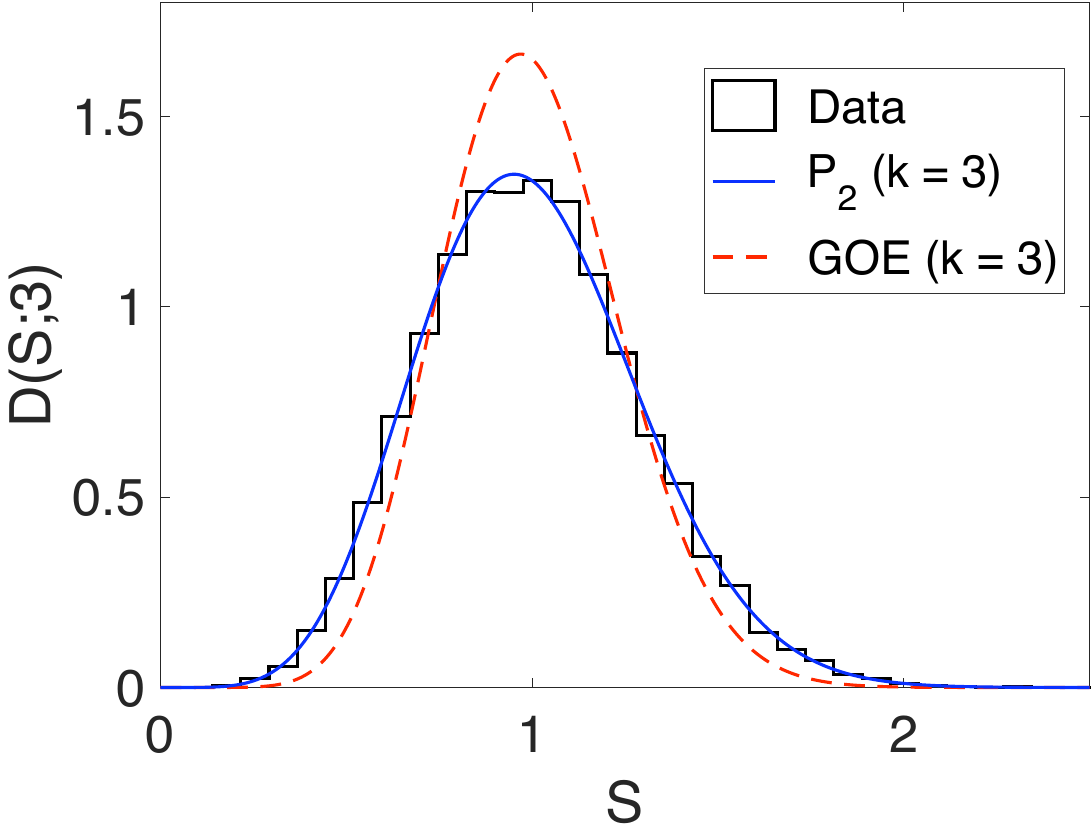}}
\caption{\label{higherorderEGs} (Top) \emph{Second}-nearest-neighbor spacing density histogram for the pseudotrajectory shown in the left panel of Fig.~\ref{skewEG}. The unbroken (blue) curve is the Ginibre distribution $P_G(S)$ [Eq.~(\ref{WigGinib})]. The dashed (red) curve is the Wigner surmise $P_W(S;k=2,\beta=1)$ [Eq.~(\ref{onetwoGOEGSEWS})]. (Bottom) \emph{Third}-nearest-neighbor spacing density histogram for the pseudotrajectory shown in the left panel of Fig.~\ref{skewEG}. The unbroken (blue) curve is the Wigner surmise $P_W(S;k=1,\beta=5)$ [Eq.~(\ref{P2Keq3})]. The dashed (red) curve is the Wigner surmise $P_W(S;k=3,\beta=1)$ [Eq.~(\ref{thirdGOE})].}
\end{figure}

Although not directly relevant to the main premise of the paper, it is worthwhile to briefly comment on and discuss higher-order spacing distributions. For all the examples considered in this paper, the higher-order spacing statistics are found to be consistent with those theoretically predicted for $\mathbf{P}_2$. Note however that the higher-order spacing statistics of $\mathbf{P}_2$ do not generally 
coincide with higher-order GOE spacing statistics. For example, the \emph{second}-NNSD 
for GOE eigenvalues is the same as the NNSD 
for eigenvalues from the Gaussian symplectic ensemble (GSE) \cite{MehtaDyson}: 
\begin{equation}\label{onetwoGOEGSE}
P_{GOE}(S;k=2)=P_{GSE}(S;k=1).
\end{equation}
The \emph{second}-NNSD 
for $\mathbf{P}_2$, on the other hand, coincides with the eigenvalue NNSD for the 
Ginibre ensemble of $2\times2$ complex non-Hermitian 
random matrices \cite{meandjohn}, which is given by \cite{Haake}:
\begin{equation}\label{WigGinib}
P_{G}(S)={3^4\pi^2\over2^7}S^3
\exp\left(-{3^2\pi\over2^4}S^2\right). 
\end{equation}
The `Wigner surmise' approximation for Eq.~(\ref{onetwoGOEGSE}) is given by (using the notation of Ref.~\cite{meandjohn})
\begin{equation}\label{onetwoGOEGSEWS}
P_W(S;k=2,\beta=1)=P_W(S;k=1,\beta=4)={2^{18}\over3^6\pi^3}S^4\exp\left(-{64\over9\pi}S^2\right).
\end{equation}
As an example, consider again the skew map of Sec.~\ref{modexample2}. For the pseudotrajectory shown in the left panel of Fig.~\ref{skewEG}, the density histogram of the (scaled) \emph{second}-nearest-neighbor spacings is shown in the top panel of Fig.~\ref{higherorderEGs}. The agreement with the Ginibre distribution [Eq.~(\ref{WigGinib})] and discordance with the Wigner surmise $P_W(S;k=2,\beta=1)$ [Eq.~(\ref{onetwoGOEGSEWS})] is evident. The result for the \emph{third}-nearest-neighbor spacings is shown in the bottom panel of Fig.~\ref{higherorderEGs}. Although not widely known, the \emph{third}-NNSD for $\mathbf{P}_2$ coincides with the Wigner surmise for the $\beta=5$ Hermite ensemble and is given by \cite{meandjohn}
\begin{equation}\label{P2Keq3}
P_W(S;k=1,\beta=5)={15^6\pi^3\over2^{24}}S^5\exp\left(-{15^2\pi\over2^8}S^2\right).
\end{equation}
(The spacing distributions of the $\beta$-Hermite ensembles are discussed in Ref.~\cite{BHEpap}.) The GOE Wigner surmise of order $k=3$ is given by \cite{meandjohn}
\begin{equation}\label{thirdGOE}
P_W(S;k=3,\beta=1)={2^{68}\over3(35)^{10}\pi^5}S^8\exp\left(-{2^{14}\over35^2\pi}S^2\right).
\end{equation}
The agreement with the \emph{third}-NNSD for $\mathbf{P}_2$ [Eq.~(\ref{P2Keq3})] and discordance with the Wigner surmise $P_W(S;k=3,\beta=1)$ [Eq.~(\ref{thirdGOE})] is again evident.

In general, as $k$ (i.e., the order of the spacing) increases, the discrepancy between the $k$th-NNSD for $\mathbf{P}_2$ and the GOE Wigner surmise of order $k$ [$P_W(S;k,\beta=1)$] increases. Note that the higher-order spacing distributions for $\mathbf{P}_2$ and the higher-order Wigner surmises for the GOE do not coincide at any common order $k>1$, but there are in fact a countably infinite number of cases in which a coincidence occurs between different order members from these two families of distributions (see Eq.~(13) of Ref.~\cite{meandjohn}) \footnote{For example, the \emph{seventh}-NNSD for $\mathbf{P}_2$ is equal to the \emph{fourth}-order Wigner surmise for the GOE.}. 

\section{A Non-Generic 2D Ergodic Map}

\begin{figure}
\scalebox{0.413}{\includegraphics*{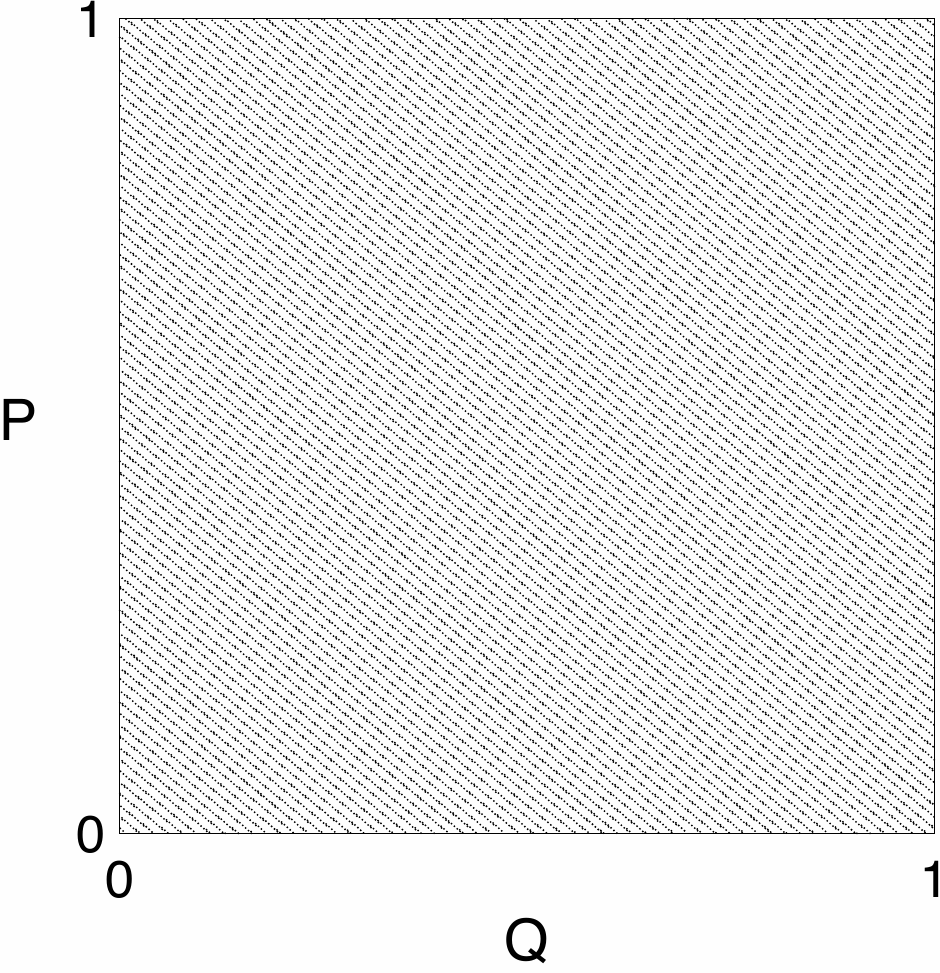}} \par \vspace*{0.75cm}
\scalebox{0.413}{\includegraphics*{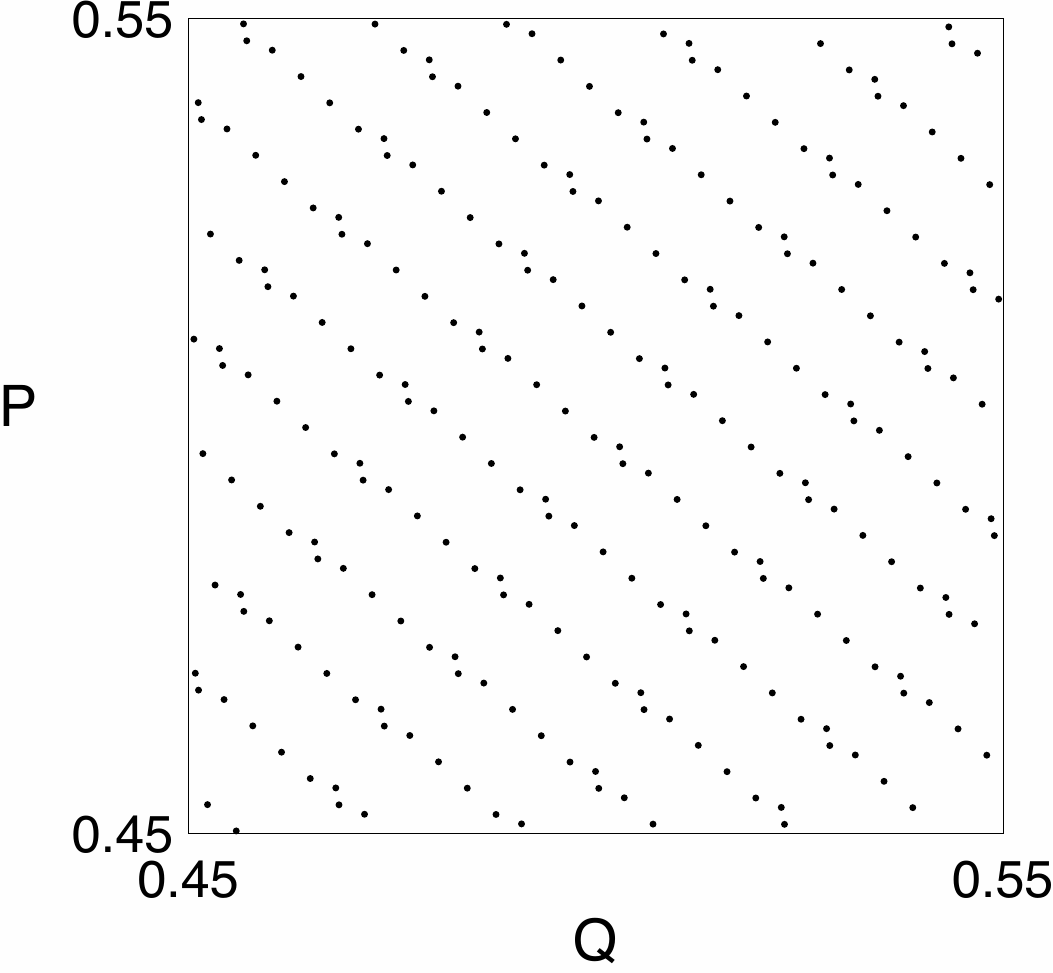}}\par \vspace*{0.75cm}
\scalebox{0.353}{\includegraphics*{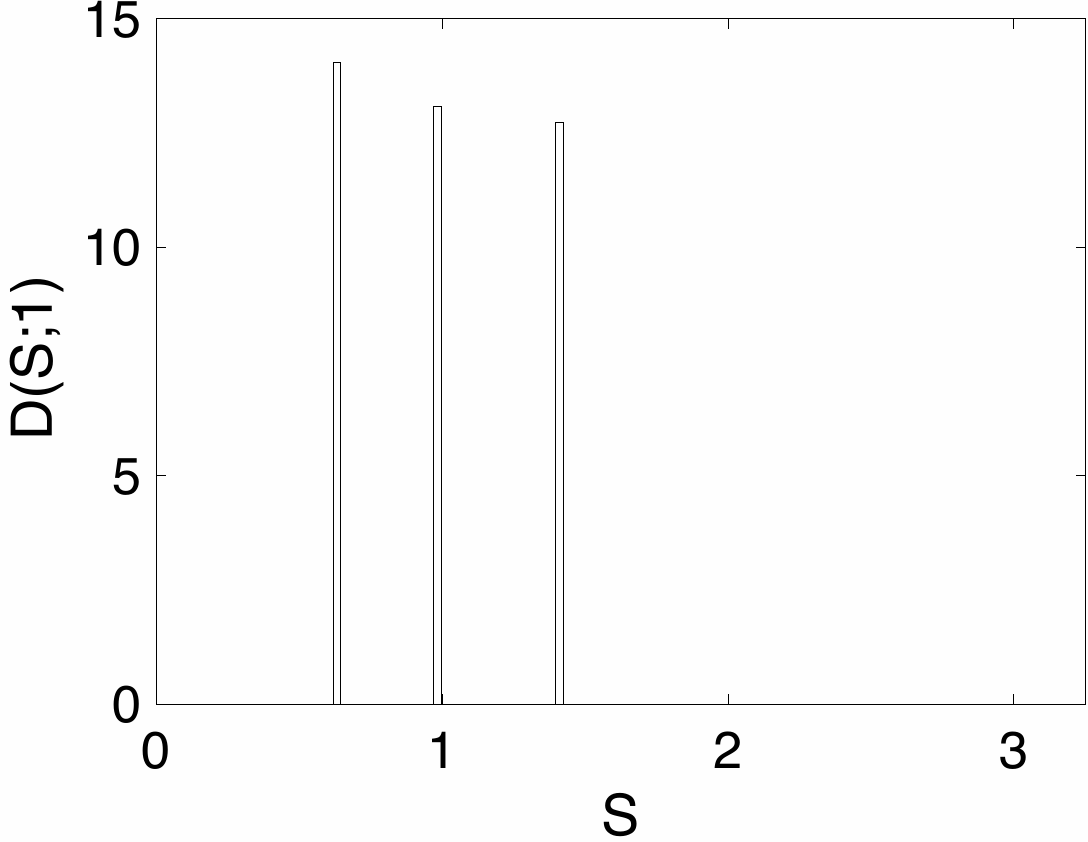}} 
\caption{\label{irrtransEG} (Top) A typical pseudotrajectory of map (\ref{irrtrans}) with $a=\sqrt{2}$ and $b=\sqrt{3}$. The pseudotrajectory was evolved from the initial point $(q_1=\sin(2/3),p_1=\cos(5/7))$, and the map was iterated $25000$ times. (Middle) A closer view of the trajectory around the region $[0.45, 0.55]\times[0.45, 0.55]$. (Bottom) Nearest-neighbor spacing density histogram for the pseudotrajectory shown in the Top panel.}
\end{figure}

The ergodic maps considered in Sec.~\ref{modexamples} can be regarded as ``generic'' 2D ergodic maps. It is expected that there will be a zero-measure set of ``non-generic'' 2D ergodic maps whose peculiarities (for example, the possession of special symmetries) render them incompatible with the current hypothesis. Irrational (incommensurate) translations or rotations on a two-torus, which are symplectic and time-reversal-invariant, are a classic example:
\begin{subequations}\label{irrtrans}
\begin{equation}\label{irrtrans1}
q_{n+1}=q_n+a,~\text{mod}~1,
\end{equation}
\begin{equation}\label{irrtrans2}
p_{n+1}=p_n+b,~\text{mod}~1,
\end{equation}
\end{subequations}
where the numbers $a$ and $b$ are such that, for any two given integers $k_1\neq0$ and $k_2\neq0$, the number $k_1a+k_2b$ is not an integer. Map (\ref{irrtrans}) is known to be ergodic (and not mixing), but its number-theoretical properties give rise to ergodic trajectories that are highly ordered. The peculiarity of these so-called quasi-periodic trajectories in the present context is that they possess a countable number of distinct interpoint spacings, and therefore cannot have a Wignerian NNSD. Figure \ref{irrtransEG} illustrates this point using map (\ref{irrtrans}) with $a=\sqrt{2}$ and $b=\sqrt{3}$. As can be discerned from the top panel of Fig.~\ref{irrtransEG}, this quasi-periodic trajectory does not have the usual random point pattern spatial structure characteristic of the examples considered in Sec.~\ref{modexamples}. Closer inspection of the trajectory (see, for example, the middle panel of Fig.~\ref{irrtransEG}) reveals that there are only \emph{three} distinguishable nearest-neighbor spacings resulting in the spiked distribution shown in the bottom panel of Fig.~\ref{irrtransEG}. The results shown in Fig.~\ref{irrtransEG} are exemplary. More generally, the number and position of the spikes will depend on the specific values of $a$ and $b$.  

\section{Discussion and Conclusion}

To summarize, the hypothesis that any typical pseudotrajectory of a 2D ergodic map will have a Wignerian nearest neighbor spacing distribution, was put forward and numerically tested. In all test cases, the hypothesis is upheld, and the range of validity of the hypothesis appears to be robust in the sense that it is not affected by the presence or absence of: (i) mixing; (ii) time-reversal symmetry; and/or (iii) dissipation. Furthermore, pseudotrajectories need not necessarily cover densely all of the available phase space in order for their NNSDs to be Wignerian in nature. As the example in Sec.~\ref{divedeg} demonstrates, pseudotrajectories evolving ergodically in any positive-measure subset of the full phase space also possess NNSDs consistent with the Wigner distribution. 

The ergodic maps considered in Sec.~\ref{modexamples} can be regarded as ``generic'' 2D ergodic maps. It is expected that there will be a zero-measure set of ``non-generic'' 2D ergodic maps whose peculiarities render them incompatible with the current hypothesis. Irrational translations or rotations on the two-torus are a classic example. The generic examples of Sec.~\ref{modexamples} represent in some sense the ideal 2D simply ergodic map. The phase space of other generic maps may possess certain features (e.g., sticky structures \cite{sticky06}) not present in the ideal examples considered here that can result in small but not insignificant deviations from a Wignerian NNSD. Identifying and understanding the effect(s) of such features on the NNSD is a problem that warrants its own study. 

It is important to stress that an ergodic map need not be chaotic. Skew maps (like the one of Sec.~\ref{modexample2}), for example, are ergodic, and while they do exhibit zero autocorrelation and a particular kind of sensitivity to initial conditions \cite{RBLOC98}, they have zero Lyapunov exponent and zero entropy \cite{Peter83}. By virtue of the last two properties as well as the fact that they have no mixing properties, skew maps are regarded as being non-chaotic. Likewise, chaotic maps are not necessarily ergodic. In fact, most of  the well-known chaotic maps \cite{JCS03} are not ergodic. Most chaotic maps do not generate trajectories that densely and uniformly cover their phase spaces (or positive-measure subsets thereof). Thus, in general, bounded aperiodic pseudotrajectories of 2D chaotic maps will not have Wignerian NNSDs. There are however a number of important special cases where such pseudotrajectories are expected to have Wignerian NNSDs. These shall be described  in a follow-up paper.

The Wigner distribution has played a fundamental role in characterizing the \emph{quantum} spectral fluctuations of classically chaotic systems. While it is interesting that the same distribution characterizes the ergodic trajectories of 2D \emph{classical} maps, there appears to be no deeper correspondence with the random matrix model of quantum chaos. To clarify this point, it is useful to refer to maps (\ref{perturbedcatTRI}) and (\ref{prtrbdmapTRSB}). When the time-reversal invariant map (\ref{perturbedcatTRI}) is quantized, the NNSD of the eigenphases is expected to be consistent with the NNSD for eigenvalues drawn from the circular orthogonal ensemble (COE), denoted here by $P_{COE}(S)$. In the limit where the matrix size goes to infinity, the COE reverts to the GOE, and hence $P_W(S)\approx P_{COE}(S)$. (Note that the sample NNSD of the eigenphases that come from quantizing map (\ref{perturbedcatTRI}) is indeed consistent with $P_W(S)$ \cite{backer03}.) On the other hand, when the non-reversible map (\ref{prtrbdmapTRSB}) is quantized, the NNSD of the eigenphases is expected to be consistent with the NNSD for eigenvalues drawn from the circular unitary ensemble (CUE), denoted here by $P_{CUE}(S)$ \cite{KM,Mthesis}. In the limit where the matrix size goes to infinity, the CUE reverts to the Gaussian unitary ensemble (GUE), and in this case the usual analytical approximation to $P_{CUE}(S)$ (the so-called Wigner surmise for the GUE) differs significantly from $P_W(S)$. The presence or absence of time-reversal symmetry thus has a significant effect on the NNSD of the quantum eigenphases. Time-reversal symmetry produces no comparable effect in the present context. Regardless of whether or not time-reversal symmetry is present, the pseudotrajectories of these 2D Anosov maps are Wignerian. It should be emphasized that the ideas and results of the present paper are not aimed at addressing the open question of \emph{why} a Wigner-\emph{like} NNSD is a common ``quantum signature of chaos'', nor should they (at present) be interpreted as being any kind of theoretical justification for this commonly observed quantum phenomenon. The goal here is merely to introduce the novel way in which the Wigner distribution enters the domain of \emph{classical} mechanics. 

The numerical results of Sec.~\ref{modexamples} describe the interpoint nearest-neighbor spacings of \emph{pseudo}trajectories. This begs the question: Do the same results hold for the \emph{exact} trajectories of a 2D ergodic  map? This is a question that is difficult to answer definitively without detailed analysis. Chaotic pseudotrajectories emulate the true dynamics of a given system only when `shadowed' by exact trajectories of the system, and are otherwise only meaningful in a statistical sense. Shadowing of numerical trajectories is a fundamental issue, in particular, for strictly non-hyperbolic chaotic systems. In hyperbolic systems, which is an extreme and rather exceptional case, the existence of shadowing trajectories for all pseudotrajectories is guaranteed (see, for example, Ref.~\cite{katok}). For such systems, conclusions about the statistical properties of numerical trajectories will also generally apply to exact trajectories. Long shadowing trajectories are not precluded for all non-hyperbolic chaotic systems (for example, the shadowing property has been established for the standard map \cite{SM1}), but their consideration introduces technical questions about how accurate and for how long numerical trajectories are valid. The maps used in Secs.~\ref{modexample1}, \ref{modexample3}, and \ref{modexample4} are all 2D Anosov maps, which are uniformly hyperbolic. Together with the appropriate shadowing theorems for hyperbolic systems, the numerical results of Secs.~\ref{modexample1}, \ref{modexample3}, and \ref{modexample4} thus motivate the following proposition: \emph{Typical trajectories of 2D Anosov maps have a Wignerian nearest-neighbor spacing distribution} \footnote{It is important to emphasize that the interpoint spacings implicitly being referred to in this statement are measured using the standard \emph{Euclidean} metric.}. 

The idea that the ergodic trajectories of 2D maps can be modeled by a homogeneous 2D Poisson point process can be generalized to higher (or lower) dimensions. For a $d$-dimensional ergodic map, the pertinent model is then the homogeneous Poisson point process in $\mathbb{R}^d$ (henceforth denoted by $\mathbf{P}_d$). When $d\neq2$, the \emph{Euclidean} NNSD for $\mathbf{P}_d$ is not given by the Wigner distribution; it is actually given by the Brody distribution \footnote{See the Introduction of Ref.~\cite{Pdspaper} for a derivation of this fact.} with the usual Brody parameter $q=(d-1)$. For a restricted set of metrics (including the Euclidean), the Wigner result is only valid when $d=2$ \cite{menonEuclid}. One can construct the NNSD for $\mathbf{P}_d$ using non-Euclidean metrics, and (for any integer value of $d$) there exist metrics for which the NNSD for $\mathbf{P}_d$ is given by the Wigner distribution \cite{menonEuclid}. Thus, armed with an appropriate non-Euclidean metric for determining the interpoint spacings, the Wigner distribution can be reinstated as the governing distribution when $d\neq2$. An analogous numerical study of the NNSD for $d$-dimensional ergodic maps ($d\neq2$) will hopefully be the subject of a future publication.

\end{document}